\documentclass[preprint,cite,epsf]{JHEP3}
\usepackage{epsfig,multicol}
\usepackage{latexsym}
\title{\Large The Low Energy Limit of the Noncommutative Wess-Zumino Model}
\author{H. O. Girotti\\Instituto de F\'\i sica, Universidade Federal do Rio Grande
do Sul\\ Caixa Postal 15051, 91501-970 - Porto Alegre, RS, Brazil\\ E-mail:
\email{hgirotti@if.ufrgs.br}}
\author{M. Gomes \\ 
Instituto de F\'\i sica, Universidade de S\~ao Paulo\\
Caixa Postal 66318, 05315-970, S\~ao Paulo - SP, Brazil\\
E-mail:\email{mgomes@fma.if.usp.br}}
\author{A. Yu. Petrov\\
Tomsk State Pedagogical University, 634041 Tomsk, Russia\\
E-mail:\email{petrov@fma.if.usp.br}; \email{petrov@tspu.edu.ru}}
\author{V. O. Rivelles\\Instituto de F\'\i sica, Universidade de S\~ao Paulo\\
 Caixa Postal 66318, 05315-970, S\~ao Paulo - SP, Brazil\\
E-mail:\email{rivelles@fma.if.usp.br}}
\author{A. J. da Silva\\  Instituto de F\'\i sica, Universidade de S\~ao Paulo\\ Caixa Postal 66318, 05315-970, S\~ao Paulo - SP, Brazil\\
E-mail:\email{ajsilva@fma.if.usp.br}}

\abstract{
The noncommutative Wess-Zumino model is used as a prototype for
studying the low energy behavior of a renormalizable noncommutative
field theory. We start by deriving the potentials mediating the
fermion-fermion and boson-boson interactions in the nonrelativistic
regime. The quantum counterparts of these potentials are afflicted by
ordering ambiguities but we show that there exists an ordering
prescription which makes them Hermitean. For space/space
noncommutativity it turns out that Majorana fermions may be pictured
as rods oriented perpendicularly to the direction of motion showing a
lack of locality, while bosons remain insensitive to the effects of
the noncommutativity. For time/space noncommutativity bosons and
fermions can be regarded as rods oriented along the direction of
motion.  For both cases of noncommutativity the scattering state
describes scattered waves, with at least one wave having negative time
delay  signalizing the underlying nonlocality. The superfield
formulation of the model is used to compute the corresponding
effective action in the one- and two-loop approximations. In the case
of time/space noncommutativity, unitarity is violated in the
relativistic regime. However, this does not preclude the existence of
a unitary low energy limit.
}

\begin{document}
%
%
%
%
%
\newcounter{multieqs}
%
\newcommand{\eqnumber}{\addtocounter{multieqs}{1}
                       \addtocounter{equation}{-1}}
%
\newcommand{\glabel}[1]
	   {\addtocounter{equation}{-1}
            \label{#1}
            \addtocounter{equation}{1}}
%
\newenvironment{eqnalpha}
{
\setcounter{multieqs}{0}
\addtocounter{equation}{1}
\renewcommand{\theequation}{\thesection.\arabic{equation}.\alph{multieqs}}
\begin{eqnarray}
}
{
\end{eqnarray}
\setcounter{multieqs}{0}
\renewcommand{\theequation}{\thesection.\arabic{equation}}
}
%
\newcommand{\begalph}
{
\setcounter{multieqs}{0}
\addtocounter{equation}{1}
\renewcommand{\theequation}{\thesection.\arabic{equation}\alph{multieqs}}
}
\newcommand{\alphend}
{
\setcounter{multieqs}{0}
\renewcommand{\theequation}{\thesection.\arabic{equation}}
}

\section{Introduction}
Noncommutative (NC) field theories present many unusual
properties. Thus, it is not surprising that many studies have been
devoted to understand the new aspects of these theories
(see \cite{Douglas,Szabo} for recent reviews).
Their non-local character gives rise to a mixing of ultraviolet (UV)
and infrared (IR) divergences which usually spoils the
renormalizability of the model \cite{Minwalla}. This peculiar property
has been investigated in the context of scalar\cite{Arefeva,Roiban,Griguolo},
gauge \cite{YM1,Benaoum,Hayakawa,Chaichian,YM2,YM3,YM4,Bonora,Rajaraman}
and supersymmetric \cite{Sheikh,Schaposnik,Gi1,Gomes,Zanon} theories.
When the noncommutativity involves the time coordinate the theory violates causality and unitarity, as has been discussed in
\cite{Susskind,Gomis1}. In particular, it was shown that the
scattering of localized quanta in NC field theory in 1+1 dimensions
can be pictured as realized by rods moving in space-time.  All these
effects are consequences of the non-local structure induced by the
noncommutativity and are so subtle that a deep understanding is highly
desirable.  On the other hand, in higher dimensions, the lack of
renormalizability induced by UV/IR mixing is quite worrisome.  Even if
one has succeeded in controlling the renormalization problem it still
remains to make sure that the aforementioned non-local effects persist in
renormalizable NC field theories \cite{Gracia,Campbell,Petriello}. The only 4D renormalizable NC field theory known at present is the Wess-Zumino model \cite{Gi1}. Hence,
we have at our disposal an appropriate model for studying the
non-local effects produced by the noncommutativity. As we will show
the main features of nonlocality are still present in the NC
Wess-Zumino model.

To study the non-local effects we consider the NC Wess-Zumino model
and determine the non-relativistic potentials mediating the
fermion-fermion and boson-boson scattering along the lines of
\cite{Sak,Girotti2}. In the case of space/space noncommutativity we
find that the potential for boson-boson scattering receives no NC
contribution. The fermion-fermion potential, however, has a NC
correction which leads to the interpretation that, in a
nonrelativistic scattering, fermionic quanta behave like rods oriented
perpendicular to their respective momenta and having lengths
proportional to the momenta strength. This extends to higher
dimensions the picture that was found in \cite{Susskind} for lower
dimensions. In the time/space NC case we find that both, boson-boson
and fermion-fermion potentials receive NC velocity dependent
corrections leading to ordering ambiguities.  These potentials can be
made Hermitean by an appropriate ordering choice for products of
noncommuting operators.  It follows afterwards that both bosons and
fermions can be viewed as rods oriented along the direction of the
momenta. The rod length, however, is constant and proportional to the
NC parameter. We also find the scattered waves and show the existence
of advanced waves which is a further manifestation of
nonlocality. Finally, we use the superfield formalism to compute,
within the relativistic regime, the one- and two-loop non-planar
corrections to the effective action. In the case of time/space NC we
find that the just mentioned contributions violate the unitarity
constraints.

The plan of this work is as follows. We start in section \ref{sec2} by
presenting the formulation of the NC WZ model in terms of field
components. In section \ref{sec21} we calculate the tree
approximations of the fermion-fermion and boson-boson elastic
scattering amplitudes, in the low energy limit. In \ref{sec22}, the
effective quantum mechanical  potentials mediating the fermion-fermion
and boson-boson interactions  are determined.  We discuss, then, the existence of an
effective Hermitean Hamiltonian acting as generator of the low energy
dynamics. Afterwards, we construct and stress the relevant features of
the scattering states in the cases of space/space and time/space
noncommutativity. In section \ref{sec3} by taking advantage of the
formulation of the model in terms of superfields we calculate the one-
and two-loop contributions to the effective action. In the case of
time/space  noncommutativity this effective action also 
exhibits unitarity violation.

\section{Tree Level Analysis\label{sec2}}
The Lagrangian density describing the dynamics of the NC WZ model is\cite{Gi1}

\begin{eqnarray}
\label{1}
{\cal L} &=&\frac12 [ A(-\partial^2)A+  B(-\partial^2)B
+  \overline \psi(i\not \! \partial - m)\psi + F^2 + G^2]\nonumber\\
&& + m F A + m G B + 
g (F\star A \star
A- F\star B \star B  \nonumber\\
&&+ G\star A \star B+ G \star B \star A -\overline \psi
\star \psi\star A - \overline \psi\star  i\gamma_5 \psi \star B),
\end{eqnarray}

\noindent
where $A$ is a scalar field, $B$ is a pseudo scalar field, $\psi$ is a
Majorana spinor field and $F$ and $G$ are, respectively, scalar and
pseudoscalar auxiliary fields. It was obtained by extending the WZ
model to a NC space. In the NC model there are neither
quadratic nor linear divergences. As a consequence, the IR/UV mixing
gives rise only to integrable logarithmic infrared divergences
\cite{Gi1,Bichl}. The Moyal ($\star$) product
obeys the rule \cite{Filk}

\begin{eqnarray}
\label{2}
\int dx \phi_1(x)\star\phi_2(x)\star...\star\phi_n(x)& =&\int \prod \frac{d^4k_i}{(2\pi)^4} (2\pi)^4\delta^{(4)}(k_1+k_2+\ldots +
k_n)\nonumber \\
&\phantom a& \tilde\phi_1(k_1) \tilde \phi_2(k_2)\ldots \tilde\phi_n(k_n)
\exp(i\sum_{i<j}k_i\wedge k_j),
\end{eqnarray}

\noindent
where $\tilde \phi_i$ is the Fourier transform of the field $\phi_i$,
the index $i$ being used to distinguish different fields. We use the
notation $a\wedge b= 1/2 a^\mu b^\nu \Theta_{\mu\nu}$. For the Feynman
rules arising from (\ref{1}) we refer the reader to Ref.\cite{Gi1}.

\FIGURE{\epsfig{file=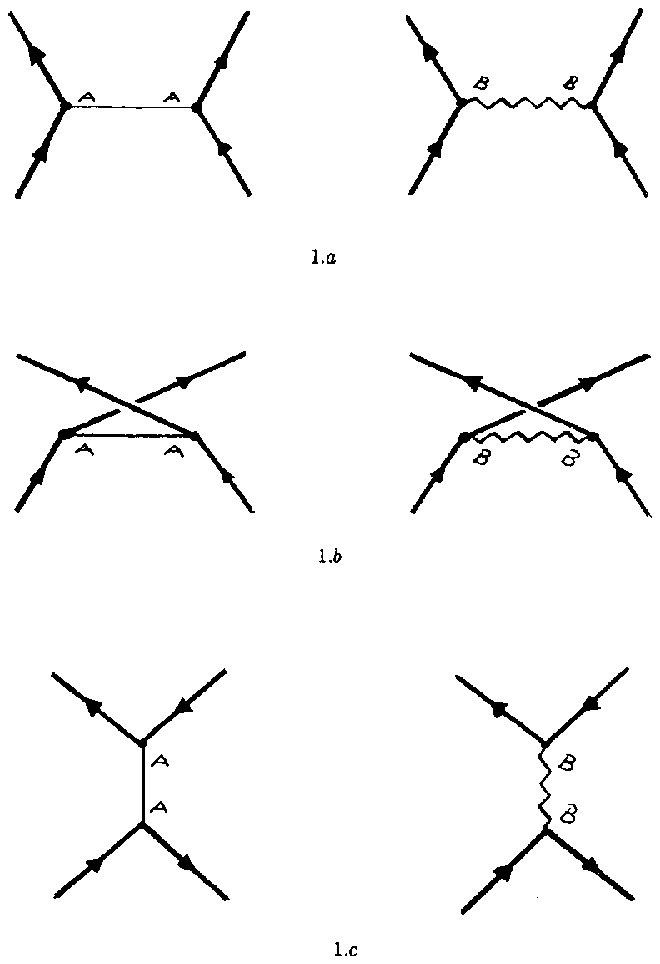,width=10cm}
\caption{{\it Lowest order graphs contributing to the scattering of two
Majorana fermions.}\label{Fig1}}}

\subsection{Tree Level Scattering\label{sec21}}
We first concentrate on the elastic scattering of two Majorana
fermions. We shall designate by $p_1, p_2$ ($p'_1, p'_2$) and by
$\epsilon_1, \epsilon_2$ ($\epsilon'_1, \epsilon'_2$) the four momenta
and z-spin components of the incoming (outgoing) particles, respectively. The
Feynman graphs contributing to this process, in the lowest order of
perturbation theory, are those depicted in Fig.\ref{Fig1}\footnote{In these
diagrams the arrows indicate the flow of fermion number rather than
momentum flow.} while the associated amplitude is given
by $R = -i(2\pi)^4 \delta^{(4)}(p'_1 + p'_2 - p_1 - p_2)\, T$, where $ T = T_a +
T_b + T_c$ and

\begalph
\begin{eqnarray}
\glabel{3}
T_a\,&=&\, K\, \cos(p'_1 \wedge p_1) \, \cos(p'_2 \wedge p_2)\,\frac{\left(F_a - F_a^5\right)}{{\cal D}_a}, \eqnumber\label{mlett:a3}\\
T_b\,&=&\,-\, K\, \cos(p'_1 \wedge p_2) \, \cos(p'_2 \wedge p_1)\,\frac{\left(F_b - F_b^5\right)}{{\cal D}_b}, \eqnumber\label{mlett:b3}\\
T_c\,&=&\, K\, \cos(p'_1 \wedge p'_2) \, \cos(p_1 \wedge p_2)\,\frac{\left(F_c - F_c^5\right)}{{\cal D}_c}.\eqnumber \label{mlett:c3}
\end{eqnarray}
\alphend

\noindent 
The correspondence between the sets of graphs $a, b, c$, in Fig.\ref{Fig1}, and the partial amplitudes $T_a, T_b, T_c$ is self explanatory. Furthermore,

\begalph
\begin{eqnarray}\glabel{4}
&&F_a \equiv \left[{\bar{u}}(\vec{p}\,'_1,\epsilon'_1) u(\vec{p}_1,\epsilon_1)\right] \left[{\bar{u}}(\vec{p}\,'_2,\epsilon'_2) u(\vec{p}_2,\epsilon_2)
\right],\eqnumber\label{mlett:a4}\\
&&F_a^5 \equiv \left[{\bar{u}}(\vec{p}\,'_1,\epsilon'_1) \gamma^5 u(\vec{p}_1,\epsilon_1)\right] \left[{\bar{u}}(\vec{p}\,'_2,\epsilon'_2) \gamma^5 u(\vec{p}_2,\epsilon_2)\right],\eqnumber\label{mlett:b4}\\
&&{\cal D}_a \equiv \left(p'_1 - p_1\right)^2 \, - \, m^2\,+\,i\epsilon, \eqnumber \label{mlett:c4}\\
&&F_b \equiv \left[{\bar{u}}(\vec{p}\,'_1,\epsilon'_1) u(\vec{p}_2,\epsilon_2)\right] \left[{\bar{u}}(\vec{p}\,'_2,\epsilon'_2) u(\vec{p}_1,\epsilon_1)\right],\eqnumber\label{mlett:d4}\\
&&F_b^5 \equiv \left[{\bar{u}}(\vec{p}\,'_1,\epsilon'_1) \gamma^5 u(\vec{p}_2,\epsilon_2)\right] \left[{\bar{u}}(\vec{p}\,'_2,\epsilon'_2) \gamma^5 u(\vec{p}_1,\epsilon_1)\right],\eqnumber\label{mlett:e4}\\
&&{\cal D}_b \equiv \left(p'_1 - p_2\right)^2 \, - \, m^2\,+\,i\epsilon,\eqnumber \label{mlett:f4}\\
&&F_c \equiv \left[{\bar{u}}(\vec{p}\,'_1,\epsilon'_1) v(\vec{p}\,'_2,\epsilon'_2)\right] \left[{\bar{v}}(\vec{p}_2,\epsilon_2) u(\vec{p}_1,\epsilon_1)
\right],\eqnumber\label{mlett:g4}\\
&&F_c^5 \equiv \left[{\bar{u}}(\vec{p}\,'_1,\epsilon'_1) \gamma^5 v(\vec{p}\,'_2,\epsilon'_2)\right] \left[{\bar{v}}(\vec{p}_2,\epsilon_2) \gamma^5 u(\vec{p}_1,\epsilon_1)\right],\eqnumber\label{mlett:h4}\\
&&{\cal D}_c \equiv \left(p_1 + p_2\right)^2 \, - \, m^2\,+\,i\epsilon, \eqnumber\label{mlett:i4}
\end{eqnarray}
\alphend

\begin{equation}
\label{5}
K\,=\, \frac{1}{\pi^2}\,\frac{g^2}{(2\pi)^4}\,\frac{m^2}{\sqrt{\omega(\vec{p}\,'_1)\omega(\vec{p}\,'_2)\omega(\vec{p}_1)\omega(\vec{p}_2)}},
\end{equation}

\noindent 
and $\omega(\vec{p}) \equiv {\sqrt{\vec{p}^{\,2} + m^2}}$. Here, the $u$'s and the $v$'s are, respectively, complete sets of positive and negative energy solutions of the free  Dirac equation. Besides orthogonality and completeness conditions they also obey

\begalph
\begin{eqnarray}\glabel{6}
&&C\,{\bar u}^T(\vec{p}, \epsilon)\,=\,v(\vec{p}, \epsilon),\eqnumber\label{mlett:a6}\\
&&C\,{\bar v}^T(\vec{p}, \epsilon)\,=\,u(\vec{p}, \epsilon),\eqnumber\label{mlett:b6}
\end{eqnarray}
\alphend

\noindent
where $C \equiv i \gamma^2 \gamma^0$ is the charge conjugation matrix and ${\bar u}^T$ (${\bar v}^T$) denotes the transpose of ${\bar u}$ (${\bar v}$). Explicit expressions for these solutions can be found in Ref.\cite{MK}.   

Now, Majorana particles and antiparticles are identical and, unlike
the case for Dirac fermions, all diagrams in Fig. \ref{Fig1}
contribute to the elastic scattering amplitude of two Majorana
quanta. Then, before going further on, we must verify that the
spin-statistics connection is at work. As expected, $T_a + T_b$
undergoes an overall change of sign when the quantum numbers of the
particles in the outgoing (or in the incoming) channel are exchanged
(see Eqs. (\ref{3}) and (\ref{4})). As for $T_c$, we notice that

\begalph

\begin{eqnarray}\glabel{7}
&&{\bar u}(p, \epsilon) v(p', \epsilon')\,=\,-\,{\bar u}(p', \epsilon') v(p, 
\epsilon), \eqnumber\label{mlett:a7}\\
&&{\bar u}(p, \epsilon) \gamma^5 v(p', \epsilon')\,=\,-\,{\bar u}(p', \epsilon') \gamma^5 v(p, \epsilon), \eqnumber\label{mlett:b7}
\end{eqnarray}
\alphend

\noindent
are just direct consequences of Eq.(\ref{6}). Thus, $T_c$, alone, also changes sign under the exchange of the outgoing (or incoming) particles and, therefore, $T_a + T_b + T_c$ is antisymmetric.

The main purpose in this paper is to disentangle the relevant features
of the low energy regime of the NC WZ model. Since noncommutativity breaks Lorentz invariance, we must carry out this task in an specific frame
of reference that we choose to be the center of mass (CM) frame. Here,
the two body kinematics becomes simpler because one has
that $p_1 = (\omega, \vec{p})$, $p_2 = (\omega, -\vec{p})$, $p'_1 = (\omega,
\vec{p}\,')$, $p'_2 = (\omega, -\vec{p}\,')$, $|\vec{p}\,'| = |\vec{p}\,|$, and
$\omega = \omega(\vec{p})$. This facilitates the calculation of all terms of the form

\begin{equation}
\label{9}
\left[ \frac{m}{\pi \omega(\vec{p})}\right]^2\,\frac{\left(F - F^5\right)}{{\cal D}}, 
\end{equation}

\noindent
in Eqs.(\ref{3}). By disregarding all contributions of order $(|\vec{p}\,|/m)^2$ and higher, and after some algebra one arrives at

\begalph
\begin{eqnarray}\glabel{11}
 T^L_a\,&=&\,-\frac{1}{(2\pi)^4}\,\left(\frac{g}{\pi \, m}\right)^2\, 
\delta_{\epsilon'_1 \epsilon_1}\, \delta_{\epsilon'_2 \epsilon_2}\,
\left[\frac{1}{2}\cos\left( m \Theta_{0j} k^j\right)\,+\,\frac{1}{2}\cos\left(p^i \Theta_{ij}
k^j\right)\right]\,,\eqnumber\label{mletta11}\\
T^L_b\,&=&\,\,\frac{1}{(2\pi)^4}\,\left(\frac{g}{\pi \, m}\right)^2\, 
\delta_{\epsilon'_1 \epsilon_2}\, \delta_{\epsilon'_2 \epsilon_1}\,
\left[\frac{1}{2}\cos\left( m \Theta_{0j} k'\,^j\right)\,+\,\frac{1}{2}\cos\left(p^i \Theta_{ij}
k'\,^j\right)\right]\,,
\eqnumber\label{mlett:b11}\\
T^L_c\,&=&\,\,\frac{1}{3(2\pi)^4}\, \left(\frac{g}{\pi \, m}\right)^2\,
\left\{\delta_{\epsilon'_1 \epsilon_1}\, \delta_{\epsilon'_2 \epsilon_2}\,
\cos\left(m \Theta_{0j} p^j\right)\cos\left[m \Theta_{0j} \left(p^j - k^j\right)\right]\right.
\nonumber\\
&&\left. -\,\delta_{\epsilon'_1 \epsilon_2}\, \delta_{\epsilon'_2 \epsilon_1}\,
\cos\left(m \Theta_{0j} p^j\right)\cos\left[m \Theta_{0j} \left(p^j-k'\,^j \right)\right]
\right\}\,,
\eqnumber\label{mlett:c11}
\end{eqnarray}
\alphend

\noindent
where $k^j \equiv p^j - p'\,^j$ ($k'\,^j \equiv p^j + p'\,^j$) denotes the momentum transferred in the direct (exchange) scattering while 
the superscript $L$ signalizes that the above expressions only hold
true for the low energy regime. It is worth mentioning 
that the dominant contributions to $ T^L_a$ and $T^L_b$ are made by those
diagrams in Fig. \ref{Fig1}$a$ and Fig.\ref{Fig1}$b$ not containing
the vertices $i \gamma^5 $, while, on the other hand, the dominant
contribution to $ T^L_c $ comes from the diagram in Fig.\ref{Fig1}$c$ with
vertices $ i \gamma^5 $. Clearly, $T^L_a + T^L_b +T^L_c $ is antisymmetric under the exchange $\epsilon_1'\leftrightarrow \epsilon'_2$, $\vec{p}\,'\rightarrow -\vec{p}\,'$ ($k^j \leftrightarrow k'\,^j$), as it must be. Also notice that, in the CM frame of reference, only the cosine factors introduced by the time/space noncommutativity are present in $T^L_c$.

\FIGURE{\epsfig{file=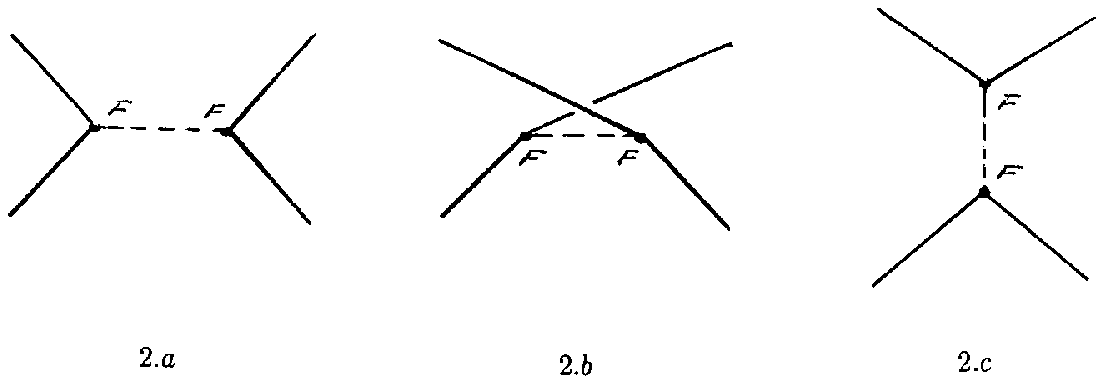,width=10cm}
\caption{{\it Lowest order graphs contributing to the scattering of two $A$-quanta.}\label{Fig2}}}

We look next for the elastic scattering amplitude involving two
$A$-field quanta. The diagrams contributing to this process, in the
lowest order of perturbation theory, are depicted in Fig.\ref{Fig2}. The
corresponding (symmetric) amplitude, already written in the CM frame of
reference, can be cast as ${\bar R} = -i(2\pi)^4 \delta^{(4)}(p'_1 + p'_2 - p_1 -
p_2)\,{\bar T}$, where $ {\bar T} = {\bar T}_a + {\bar T}_b + {\bar
T}_c$ and

\begalph
\begin{eqnarray}\glabel{12}
{\bar T}_a \,&=&\,\,\frac{g^2}{(2\pi)^4}\,\left(\frac{1}{2\pi 
\omega(\vec{p})}\right)^2\,\left[\frac{1}{2}\cos\left( m \Theta_{0j} k^j\right)\,+\,\frac{1}{2}\cos\left(p^i \Theta_{ij}
k^j\right)\right]\, \frac{1}{\bar {\cal D}}_a,\eqnumber \label{mlett:a12}\\
{\bar T}_b \,&=&\,\,\frac{g^2}{(2\pi)^4}\,\left(\frac{1}{2\pi 
\omega(\vec{p})}\right)^2\,\left[\frac{1}{2}\cos\left( m \Theta_{0j} k'\,^j\right)\,+\,\frac{1}{2}\cos\left(p^i \Theta_{ij}
k'\,^j\right)\right]\,\frac{1}{\bar {\cal D}}_b, \eqnumber\label{mlett:b12}\\
{\bar T}_c \,&=&\,\,\frac{g^2}{2 (2\pi)^4}\,\left(\frac{1}{2\pi 
\omega(\vec{p})}\right)^2\,\left\{\cos\left(m \Theta_{0j} p^j\right)\cos\left[m \Theta_{0j} \left(p^j - k^j\right)\right]\right.\nonumber\\
&&\left. +
\cos\left(m \Theta_{0j} p^j\right)\cos\left[m \Theta_{0j} \left (p^j-k'\,^j \right)\right]\right\} \,\frac{1}{\bar {\cal D}}_c\,. \eqnumber\label{mlett:c12}
\end{eqnarray}
\alphend            

\noindent
As far as the low energy limit is concerned, the main difference
between the fermionic and bosonic scattering processes rests, roughly speaking,
on the structure of the propagators mediating the interaction. Indeed,
the propagators involved in the fermionic amplitude are those of the
fields $A$ and $B$, namely\cite{Gi1},

\begin{equation}
\label{13}
\Delta_{AA}(p)\,=\,\Delta_{BB}(p)\,=\,i\,{\cal D}^{-1}(p)\,=\,\frac{i}{p^2 - m^2 
+ i\epsilon},
\end{equation}

\noindent
which, in all the three cases (a, b, and c), yield a nonvanishing
contribution at low energies (see Eqs.(\ref{mlett:c4}), (\ref{mlett:f4}) and (\ref{mlett:i4})).
On the other hand, the propagator involved in the bosonic amplitude is that of the $F$-field, i.e.\cite{Gi1},

\begin{equation}
\label{14}
\Delta_{FF}\,=\,i\,{\bar {\cal D}}^{-1}(p)\,=\,i\, \frac{p^2}{p^2 - m^2 +i\epsilon},
\end{equation}

\noindent
which in turns implies that

\begalph
\begin{eqnarray}\glabel{15}
&&{\bar {\cal D}}_a^{-1}\,=\,\frac{2\,\biggl|\frac{\vec{p}}{m}\biggr|^2\,\left(1 - \cos \theta \right)}{ 1 + 2\,\biggl|\frac{\vec{p}}{m}\biggr|^2\,\left(1 - \cos \theta \right)}\,=\,{\cal O}\left( \biggl|\frac{\vec{p}}{m}\biggr|^2\right),\eqnumber
\label{mlett:a15}\\
&&{\bar {\cal D}}_b^{-1}\,=\,\frac{2\,\biggl|\frac{\vec{p}}{m}\biggr|^2\,\left(1 + \cos \theta \right)}{ 1 + 2\,\biggl|\frac{\vec{p}}{m}\biggr|^2\,\left(1 + \cos \theta \right)}\,=\,{\cal O}\left( \biggl|\frac{\vec{p}}{m}
\biggr|^2\right),\eqnumber\label{mlett:b15}\\
&&{\bar {\cal D}}_c^{-1}\,=\,\frac{4\, +\, 4\,\biggl|\frac{\vec{p}}{m}\biggr|^2\,}{ 3 + 4\,\biggl|\frac{\vec{p}}{m}\biggr|^2}\,=\,\frac{4}{3 }\,\left[1 + {\cal O}\left( \biggl|\frac{\vec{p}}{m}\biggr|^2\right)\right].\eqnumber\label{mlett:c15} 
\end{eqnarray}
\alphend

\noindent
Therefore, at the limit where all the contributions of order $(|\vec{p}\,|/m)^2$ become neglectable, the amplitudes ${\bar T}_a$ and ${\bar T}_b$ vanish whereas ${\bar T}_c$ survives and is found to read

\begin{equation}
\label{16}
{\bar T}^L_c =  \frac{1}{6 (2\pi)^4} \left(\frac{g}{\pi m}\right)^2
\cos\left(m \Theta_{0j} p^j\right)\left\{\cos\left[m \Theta_{0j} \left(p^j - k^j\right)\right]
+ \cos\left[m \Theta_{0j} \left(p^j-k'\,^j \right)\right]\right\}.
\end{equation}

\subsection{The Effective Quantum Mechanical Potential\label{sec22}}
We shall next start thinking of the amplitudes in Eqs.(\ref{11}) and
(\ref{16}) as of scattering amplitudes deriving from a set of
potentials. These potentials are defined as the Fourier transforms (FT),
with respect to the transferred momentum ($\vec{k}$), of the
respective direct scattering amplitudes. This is due to the fact that
the use, in nonrelativistic quantum mechanics, of antisymmetric wave
functions for fermions and of symmetric wave functions for bosons
automatically takes care of the contributions due to exchange
scattering\cite{Sak}.  Whenever the amplitudes depend only on 
$\vec k$  the corresponding FT  will be
local, depending only on a relative coordinate $\vec r$. However, if,
as it happens here, the amplitudes depend  not only on $\vec k$ but also on the initial momentum of the scattered
particle $(\vec p)$, the FT will be a function of both $\vec r$
and  $\vec p$.  As the  momentum and position operators do
not commute the construction of  potential operators from these
FT may be jeopardized by ordering problems. In that
situation, we will proceed as follows: In the FT of the
amplitudes we promote the relative coordinate and momentum to
noncommuting canonical conjugated variables and then solve possible
ordering ambiguities by requiring hermiticity of the resulting
expression. A posteriori, we shall verify that this is in fact an effective
potential in the sense that its momentum space matrix elements
correctly reproduce the scattering amplitudes  that we had at the
very  start of this construction.

We are, therefore, led to introduce

\begin{equation}
\label{17}
\delta_{\epsilon'_1 \epsilon_1}\, \delta_{\epsilon'_2 \epsilon_2}\,{\cal M}^F(\vec{k}, \vec{p})\,\equiv\,T_a^L(\vec{k}, \vec{p})\,+\,T_{c,dir}^L(\vec{k}, \vec{p})\,
\end{equation}

\noindent
and

\begin{equation}
\label{18}
{\cal M}^B(\vec{k},\vec{p})\,\equiv \,{\bar T}_{c,dir}^L(\vec{k}, \vec{p})\,,
\end{equation}

\noindent
in terms of which the desired FT ($V^F$ and $V^B$) are given by

\begin{equation}
\label{19}
V^{F,B}({\vec r}, \vec{p})\,=\,(2 \pi)^3\,\int d^3k\,{\cal M}^{F,B}(\vec{k}, \vec{p})\,e^{i \vec{k} \cdot {\vec r}}.
\end{equation}

\noindent
In the equations above, the superscripts $F$ and $B$ identify,
respectively, the fermionic and bosonic amplitudes and
Fourier transforms. Also, the subscript $dir$ specifies that only the direct
pieces of the amplitudes $T_c^L$ and ${\bar T}_c^L$ enter in the
calculation of the respective ${\cal M}$. Once $V^{F,B}({\vec r},
\vec{p})$ have been found one has to look for their corresponding quantum operators,
${\hat V}^{F,B}\,({\vec R}, {\vec P})$, by performing the replacements
${\vec r} \rightarrow {\vec R}, {\vec p} \rightarrow {\vec P}$, where
$\vec{R}$ and $\vec{P}$ are the Cartesian position and momentum
operators obeying, by assumption, the canonical commutation relations
$\left[R^j , R^l\right] = \left[P^j , P^l\right]= 0$ and $\left[R^j ,
P^l\right]= i\,\delta^{jl}$. By putting all this together one is led
to the Hermitean forms

\begin{eqnarray}
\label{29}
{\hat V}^F({\vec R}, {\vec P})\,&=&\,-\,\left(\frac{g}{m}\right)^2\,\int \frac{d^3k}{(2 \pi)^3}
\left( e^{ i k^l R^l } \, e^{i k^l \Theta_{lj} P^j } +  e^{i k^l R^l}\, e^{- i k^l \Theta_{lj} P^j} \right)\nonumber\\
\,&-&\,\frac{2}{3}\left(\frac{g}{m}\right)^2\,\left[\delta^{(3)}\left({\vec R} + m {\vec \Theta}\right)\,+\,\delta^{(3)}\left({\vec R} - m {\vec \Theta}\right)\right]\nonumber\\
&+&\,\frac{1}{3}\left(\frac{g}{m}\right)^2\,\left[\,
\delta^{(3)}\left({\vec R} - m {\vec \Theta}\right)\,e^{- 2 i m {\vec \Theta} \cdot {\vec P}}
\,+\,e^{2 i m {\vec \Theta} \cdot {\vec P}}\,\delta^{(3)}\left({\vec R} - m {\vec \Theta}\right)\right]\,,
\end{eqnarray}

\begin{eqnarray}
\label{30}
{\hat V}^B({\vec R}, {\vec P})\,&=&\,\frac{1}{6}\left(\frac{g}{m}\right)^2\,
\left[\delta^{(3)}\left({\vec R} + m {\vec \Theta}\right)\,+\,\delta^{(3)}
\left({\vec R} - m {\vec \Theta}\right)\right]\nonumber\\
&+&\,\frac{1}{6}\left(\frac{g}{m}\right)^2\,\left[
\delta^{(3)}\left({\vec R} - m {\vec \Theta}\right)\,e^{- 2i m {\vec \Theta} \cdot {\vec P}}
\,+\,e^{2 i m {\vec \Theta} \cdot {\vec P}}\,\delta^{(3)}\left({\vec R} - m {\vec \Theta}\right)\right]\,,
\end{eqnarray}

\noindent
where ${\vec \Theta} \equiv \{\Theta^{0j}, j=1,2,3\}$. Notice that the
magnetic components of $\Theta_{\mu \nu}$, namely $\Theta_{ij}$, only
contribute to ${\hat V}^F$ and that such contribution is free of
ordering ambiguities, since

\begin{equation}
\label{22}
\left[k^l R^l\,,\,k^m \Theta_{mj} P^j\right]\,=\,i\,k^l \,k^m \,\Theta_{mj}\,\delta^{lj}\,=\,0,
\end{equation}

\noindent
in view of the antisymmetry of $\Theta_{mj}$. On the other hand, the
contributions to ${\hat V}^F$ and ${\hat V}^B$ originating in the
electric components of $\Theta_{\mu \nu}$, namely $\Theta_{0j}$, are
afflicted by ordering ambiguities. The relevant point is that there
exist a preferred ordering that makes ${\hat V}^{F}$ and ${\hat
V}^{B}$ both Hermitean, for arbitrary $\Theta_{\mu \nu}$. Equivalent
forms to those presented in Eqs.(\ref{29}) and (\ref{30}) can be
obtained by using

\begin{equation}
\label{222}
\delta^{(3)} \left({\vec R} - m {\vec \Theta}\right)\,\exp 
\left(- 2im {\vec \Theta} \cdot {\vec P}\right)\,=\,\exp \left(- 2im {\vec 
\Theta} \cdot {\vec P}\right)\,\delta^{(3)} \left({\vec R} + m {\vec 
\Theta}\right)\,.
\end{equation}

We shall shortly verify that the matrix elements of the operators
(\ref{29}) and (\ref{30}) agree with the original scattering amplitudes.
Before that, however, we want to make some observations about
physical aspects of these operators.

We will consider, separately, the cases of space/space ($\Theta_{0j}
=0$) and time/space ($\Theta_{ij} = 0$) noncommutativity. Hence, we
first set $\Theta_{0j} = 0$ in Eqs.(\ref{29}) and (\ref{30}). As can
be seen, the potential ${\hat V}^B$, mediating the interaction of two
$A$ quanta, remains as in the commutative case, i.e., proportional to
a delta function of the relative distance between them. The same
conclusion applies, of course, to the elastic scattering of two $B$
quanta. In short, taking the nonrelativistic limit also implies in
wiping out all the modifications induced by the space/space
noncommutativity on the bosonic scattering amplitudes. On the
contrary, Majorana fermions are sensitive to the presence of
space/space noncommutativity. Indeed, from Eq.(\ref{29}) follows that
${\hat V}^F$ can be split into planar (${\hat V}^F_P$) and nonplanar
(${\hat V}^F_{NP}$) parts depending on whether or not they depend on
$\Theta_{ij}$, i.e.,

\begin{equation}
\label{20}
{\hat V}^F({\vec R}, {\vec P)}\,=\,{\hat V}^F_P(\vec{R}, {\vec P})\,+
\,{\hat V}^F_{NP}(\vec{R}, {\vec P})\,,
\end{equation}

\noindent
with

\begalph
\begin{eqnarray}\glabel{21}
{\hat V}^F_P(\vec{R})\,&=&\,-\,\frac{{2}}{3}\,\left(\frac{g}{m}\right)^2\,\delta^{(3)} (\vec{R})\,,\eqnumber\label{mlett:a21}\\
{\hat V}^F_{NP}(\vec{R}, \vec{P})\,&=&\,-\,\left(\frac{g}{m} \right)^2 \int \, \frac{d^3k}{(2 \pi)^3} \left[ \exp \left(i k^l R^l \right) \, \exp \left(i k^l \Theta_{lj} P^j \right)\right.\nonumber\\
&& \left. +  \exp \left(i k^l R^l \right)\, \exp \left(- i k^l \Theta_{lj} P^j \right) \right]\,.\eqnumber\label{mlett:b21}
\end{eqnarray}
\alphend

\noindent
For further use in the Schr\"odinger equation, we shall be needing the position representation of ${\hat{V}^F}\left(\vec{R}, \vec{P}\right)$. >From (\ref{mlett:a21}) one easily sees that $<\vec{r}\,|{\hat{V}^F}_P|\vec {r}\,'> = - {2}/3 \,(g/ m)^2\,\delta^{(3)}(\vec{r})\,
\delta^{(3)}(\vec{r} - \vec{r}\,')$. On the other hand, for the computation of
$<\vec{r}\,|{\hat {V}^F}_{NP}|\vec {r}\,'>$ it will prove convenient to introduce the realization of $\Theta_{ij}$ in terms of the magnetic field $\vec{B}$, i.e.,

\begin{equation}
\label{24}
\Theta_{ij}\,=\,-\,\epsilon_{ijk}\,B^k\,,
\end{equation}

\noindent
where $\epsilon_{ijk}$ is the fully antisymmetric Levi-Civita tensor ($\epsilon^{123} = +1$). After straightforward calculations one arrives at

\begin{equation}
\label{27}
<\vec{r}\,|{\hat {V}^F}_{NP}|\vec {r}\,'>\,=\,- \frac{2}{(2 \pi)^2}\,\left( \frac{g}{ m}\right)^2\,\frac{1}{B^2}\,\delta^{(1)}({\vec r}_{\parallel})\,\delta^{(1)}({\vec r}_{\parallel} - {\vec r}\,'_{\parallel})\,\cos\left[\frac{\left({\vec r}_{\perp} \times {\vec r}\,'_{\perp}\right) \cdot {\vec B}}{B^2}\right]\,.
\end{equation}

\noindent
Here, ${\vec r}_{\parallel}$ (${\vec r}_{\perp}$) denotes the component of ${\vec r}$ parallel (perpendicular) to ${\vec B}$, i.e., ${\vec r}_{\parallel} = ({\vec r} \cdot {\vec B}){\vec B}/B^2$ (${\vec r}_{\perp} = - ({\vec r} \times {\vec B})\times {\vec B}/B^2$). We remark that  the momentum space matrix element

\begin{eqnarray}
\label{28}
&&<\vec{p}\,'|{\hat V}^F_{NP}|\vec{p} >\,=\,\int d^3r \int d^3r' <\vec{p}\,'|{\vec r}><{\vec r}\,|{\hat V}^F_{NP}|{\vec r}\,'><{\vec r}\,'|\vec{p} >\nonumber\\
&=&\,-\,\frac{1}{(2\pi)^3}\,\left(\frac{g}{m B}\right)^2\,\int d^2r_{\perp}\,
\exp\left(-i {\vec p}\,'_{\perp} \cdot {\vec r}_{\perp}\right) \nonumber\\
&\times&\left\{\delta^{(2)}\left[{\vec p}_{\perp} - \left({\vec r}_{\perp} \times \frac{{\vec B}}{B^2}\right)\right]\,+\,\delta^{(2)}\left[{\vec p}_{\perp} + \left({\vec r}_{\perp} \times \frac{{\vec B}}{B^2}\right)\right]\right\}\nonumber\\
&=&-\,\frac{1}{4 \pi^3}\,\left(\frac{g}{m}\right)^2\,\cos \left[\left({\vec p}_{\perp} \times {\vec p}\,'_{\perp}\right) \cdot {\vec B}\right]\,
\end{eqnarray}

\noindent
agrees with the last term in (\ref{mletta11}), as it should. We also observe that the interaction only takes place at ${\vec r}_{\perp} = \pm {\vec B}\times {\vec p}_{\perp}$. This implies that ${\vec r}_{\perp}$ must also be orthogonal to ${\vec p}_{\perp}$. Hence, in the case of space/space noncommutativity fermions may be pictured as rods oriented perpendicular to the direction of the incoming momentum. Furthermore, the right hand side of this last equation vanishes if either ${\vec p}_{\perp} \times {\vec p}\,'_{\perp} = 0$, or $\left({\vec p}_{\perp} \times {\vec p}\,'_{\perp}\right) \cdot {\vec B} = 0$, or $\vec{p} = \vec{p}_{\parallel}$, or $\vec{p}\,' = \vec{p}\,'_{\parallel}$.

In the Born approximation, the fermion-fermion elastic scattering amplitude ($f^F({\vec p}\,',{\vec p})$) can be computed at once, since $f^F({\vec p}\,',{\vec p}) = -4 \pi^2 m <\vec{p}\,'|{\hat V}^F|\vec{p}>$. In turns, the corresponding outgoing scattering state ($\Phi^{F(+)}_{\vec{p}}(\vec r)$) is found to behave asymptotically ($r \rightarrow \infty$) as follows

\begin{eqnarray}
\label{282}
&& e^{-iEt}\, \Phi^{F(+)}_{\vec{p}}(\vec r)\,\sim\,\left(\frac{1}{2\pi}\right)^{\frac{3}{2}}\left[ e^{-i \left(E t -\vec{p} \cdot {\vec r}\right)}\,+\,\frac{e^{-i\left(E t -pr\right)}}{r}\,f^F({\vec p}\,',{\vec p})\right]\nonumber\\
& \sim & \left(\frac{1}{2\pi}\right)^{\frac{3}{2}}\left\{e^{-i \left(E t - \vec{p} \cdot {\vec r}\,\right)}+\,\frac{g^2}{3\pi m} \frac{e^{- i(Et - pr)}}{r} \right.\nonumber\\
&&\left. + \frac{g^2}{2\pi m}\,\left [\frac{e^{- i\left[Et - \left(\vec{p}_{\perp} \times \vec{p}\,'_{\perp}\right) \cdot {\vec B} - pr\right]}}{r}\,+\,\frac{e^{- i\left[Et + \left(\vec{p}_{\perp} \times \vec{p}\,'_{\perp}\right) \cdot {\vec B} - pr\right]}}{r}\right]\right\}\,,
\end{eqnarray}

\noindent
where $E = \vec{p}^2/2m$ is the energy of the incoming particle. The right hand side of Eq.(\ref{282}) contains three scattered waves. The one induced by the planar part of the potential (${\hat V}^F_P$) presents no time delay. The other two originate in the nonplanar part of the potential (${\hat V}^F_{NP}$) and exhibit time delays of opposite signs and proportional to $\left(\vec{p}_{\perp} \times \vec{p}\,'_{\perp}\right) \cdot {\vec B}$. For instance, for ${\vec B}$ and ${\vec p}$ along the positive Cartesian semiaxis $x^1$ and $x^3$, respectively, one has that   
$\left(\vec{p}_{\perp} \times \vec{p}\,'_{\perp}\right) \cdot {\vec B}\,=\,- 2mEB \sin\theta \sin\phi$, were, $\theta$ and $\phi$ are the scattering and azimuthal angles, respectively. The $\phi$-dependence reflects the breaking of rotational invariance. 

We set next $\Theta_{ij} = 0$, in Eqs(\ref{29}) and (\ref{30}), and turn into analyzing the case of time/space noncommutativity. The effective potentials are now 

\begin{eqnarray}
\label{283}
{\hat {\tilde V}}^F({\vec R}, {\vec P})\,&=&\,-\,\frac{2}{3}\left(\frac{g}{m}\right)^2\,\left[\delta^{(3)}\left({\vec R} + m {\vec \Theta}\right)\,+\,\delta^{(3)}\left({\vec R} - m {\vec \Theta}\right)\right]\nonumber\\
&+&\,\frac{1}{3}\left(\frac{g}{m}\right)^2\,\left[\,
\delta^{(3)}\left({\vec R} - m {\vec \Theta}\right)\,e^{- 2 i m {\vec \Theta} \cdot {\vec P}}
\,+\,e^{2 i m {\vec \Theta} \cdot {\vec P}}\,\delta^{(3)}\left({\vec R} - m {\vec \Theta}\right)\right]\,,
\end{eqnarray}

\begin{eqnarray}
\label{284}
{\hat {\tilde V}}^B({\vec R}, {\vec P})\,&=&\,\frac{1}{6}\left(\frac{g}{m}\right)^2\,\left[\delta^{(3)}\left({\vec R} + m {\vec \Theta}\right)\,+\,\delta^{(3)}\left({\vec R} - m {\vec \Theta}\right)\right]\nonumber\\
&+&\,\frac{1}{6}\left(\frac{g}{m}\right)^2\,\left[
\delta^{(3)}\left({\vec R} - m {\vec \Theta}\right)\,e^{- i 2m {\vec \Theta} \cdot {\vec P}}
\,+\,e^{ i 2m {\vec \Theta} \cdot {\vec P}}\,\delta^{(3)}\left({\vec R} - m {\vec \Theta}\right)\right]\,,
\end{eqnarray}

\noindent
where the slight change in notation (${\hat V} \rightarrow {\hat {\tilde V}}$) is for avoiding confusion with the previous case. As before, we look first for the fermionic and bosonic elastic scattering amplitudes and then construct the asymptotic expressions for the corresponding scattering states. Analogously to (\ref{28}) and (\ref{282}) we find that

\begin{eqnarray}
\label{285}
<\vec{p}\,'|{\hat {\tilde V}}^F_{NP}|\vec{p} > &=& 
-\frac{1}{12  \pi^3}\,\left( \frac{g}{m}\right )^2\left\{2 \cos \left[m {\vec \Theta} \cdot \left( \vec{p} - \vec{p}\,'\right)\right]\,-\,\cos \left[m {\vec \Theta} \cdot \left( \vec{p} + \vec{p}\,'\right)\right]\right\}\,
\end{eqnarray}

\noindent
and

\begin{eqnarray}
\label{286}
e^{-iEt}\,{\tilde \Phi}_{\vec{p}}^{F (+)}{(\vec r)}\,&=&\,\left(\frac{1}{2\pi}\right)^{\frac{3}{2}}
\left \{e^{-i\left(Et - \vec{p} \cdot {\vec r}\right)}\right.\nonumber\\
&& \left.+\,\frac{g^2}{3 m \pi r}\,\left [{e^{-i\left[Et - m {\vec \Theta} \cdot \left( \vec{p} - \vec{p}\,'\right) - pr \right]}}\,+\,  
{e^{-i\left[Et + m {\vec \Theta} \cdot \left( \vec{p} - \vec{p}\,'\right) - pr \right]}}\right ]\right. \nonumber\\
&& \left. -\,\frac{g^2}{6 m \pi r}\,\left [{e^{-i\left[Et - m {\vec \Theta} \cdot \left( \vec{p} + \vec{p}\,'\right) - pr \right]}}\,+\,  
{e^{-i\left[Et + m {\vec \Theta} \cdot \left( \vec{p} + \vec{p}\,'\right) - pr \right]}}\right ]\right\}\,.
\end{eqnarray}

\noindent
in accordance with the calculations of the section 2.
As for the bosons, the potential in Eq.(\ref{284}) leads to

\begin{eqnarray}
\label{287}
<\vec{p}\,'|{\hat {\tilde V}}^B_{NP}|\vec{p} >\, &=&\frac{1}{24  \pi^3}\,\left( \frac{g}{m}\right )^2\,\,\left\{ \cos \left[m {\vec \Theta} \cdot \left( \vec{p} - \vec{p}\,'\right)\right]\,+\,\cos \left[m {\vec \Theta} \cdot \left( \vec{p} + \vec{p}\,'\right)\right]\right\}\,
\end{eqnarray}

\noindent
and

\begin{eqnarray}
\label{288}
e^{-iEt}\,{\tilde \Phi}_{\vec{p}}^{B (+)}{(\vec r)}\,&=&\,\left(\frac{1}{2\pi}\right)^{\frac{3}{2}}
\Bigl \{e^{-i\left(Et - \vec{p} \cdot {\vec r}\right)}\Bigr.\nonumber\\
&& \left.-\,\frac{g^2}{12 m \pi r}\,\left [{e^{-i\left[Et - m {\vec \Theta} \cdot \left( \vec{p} - \vec{p}\,'\right) - pr \right]}}\,+\,  
{e^{-i\left[Et + m {\vec \Theta} \cdot \left( \vec{p} - \vec{p}\,'\right) - pr \right]}}
\right. \right. \nonumber\\[12pt]
&& \left. \left. +\,{e^{-i\left[Et - m {\vec \Theta} \cdot \left( \vec{p} + \vec{p}\,'\right) - pr \right]}}\,+\,  
{e^{-i\left[Et + m {\vec \Theta} \cdot \left( \vec{p} + \vec{p}\,'\right) - pr \right]}}\right ]\right \}\,.
\end{eqnarray}

We stress that, presently, the interaction only takes place at ${\vec
r} = \pm (\vec{p} - \vec{p}\,')/m^2$ and ${\vec r} = \pm (\vec{p} + \vec{p}\,')/m^2$
(see Eqs.(\ref{285}) and (\ref{287})). As consequence, particles in
the forward and backward directions behave as rigid rods oriented
along the direction of the incoming momentum $\vec{p}$.  Furthermore, each
scattering state (see Eqs.(\ref{286}) and (\ref{288})) describes four
scattered waves. Two of these waves are advanced, in the sense that
the corresponding time delay is negative, analogously to what was
found in \cite{Susskind}.

\section{One and Two Loop Corrections\label{sec3}}
Our study of the low energy limit of the noncommutative WZ model
ends here. The main conclusion is that the quantum mechanics
originating in this limit is always unitary. This is not in conflict
with the existence of scattered advanced waves. Of
course, this picture may change if loop contributions are taken into
account. To see whether that really happens we shall employ the
superfield approach, which is more appropriate for calculations
involving higher orders in perturbation theory\footnote{ See for
instance Ref. \cite{Buchbinder}.}. This formulation has already been
used to find the leading contributions to the effective action in one
and two-loop orders in the case of the commutative WZ
model \cite{Buchbinder,Petrov1}.

The superfield action for the NC WZ model is \cite{Bichl}

\begin{equation}
\label{31}
S\,=\,\int d^8 z\, \bar{\Phi}\Phi\,-\,\left[\int d^6 z \,\left(\frac{1}{2}m\Phi^2\,+\,\frac{g}{3!}\Phi*\Phi*\Phi\right)\,+\,h.c.\right]\,.
\end{equation}

\noindent
Here, $\Phi$ is a chiral superfield (for its component expansion see, for instance, Ref. \cite{Buchbinder}). Moreover, the Moyal product for superfields is defined as in Eq.(\ref{2}). Notice that the noncommutativity does not involve the Grassmann coordinates. The propagators look as follows \cite{Buchbinder,Bichl}

\begalph
\begin{eqnarray}\glabel{32}
< \Phi(z_1) {\bar \Phi}(z_2) >\,&=&\,\frac{-1}{\Box + m^2}\, \delta^{(8)} (z_1 - z_2)\,,\eqnumber\label{mlett:a32}\\
< \Phi(z_1) \Phi(z_2) >\,&=&\,\frac{m}{4 \Box}\,\frac{{D}^2}{\Box + m^2}\, \delta^{(8)} (z_1 - z_2)\,,\eqnumber\label{mlett:b32}
\end{eqnarray}
\alphend

\noindent
where the $ D$ factors are associated with vertices just by the same rules as in the commutative case. A chiral vertex, with $n$ external lines, carries $(2 - n)$  factors $(-1/4){\bar D}^2$. 
In a similar way, an antichiral vertex carries $(2 - n)$ factors $(-1/4) {D}^2$. Furthermore, in momentum representation, any vertex also includes the factor $\cos (p_1 \wedge p_2)$, where $p_1$ and $p_2$ are two out of the three incoming momenta \cite{Bichl}. Just for comparison purposes we mention that the low energy direct scattering amplitudes associated with the supergraphs whose corresponding fermion component graphs are those given in Figs.\ref{Fig1}$a$, and \ref{Fig1}$c$ read, respectively,

\begalph
\begin{eqnarray}\glabel{33}
T^S_a\,&=&\,\left( \frac{g}{m}\right)^2\,\left(\frac{i}{\pi^2}\right)\,\frac{1}{(2 \pi)^4}\,\int d^4 \theta\,\,\Phi(m, \vec{p}, \theta)\,{\bar \Phi}(m,-\vec{p}, \theta)
\nonumber\\
&&\Phi(m, \vec{p}\,', \theta)\,{\bar \Phi}(m, -\vec{p}\,', \theta)
\left[\frac{1}{2}\cos\left( m \Theta_{0j} k^j\right)\,+\,\frac{1}{2}\cos\left(p^i \Theta_{ij}
k^j\right)\right]
\,,\eqnumber\label{mlett:a33}\\
T^S_c\,&=&\,-\,\frac{1}{3}\,\left(\frac{g}{m}\right)^2\,
\left(\frac{i}{\pi^2}\right)\,\frac{1}{(2 \pi)^4}\,\int d^4 \theta\,\,
\Phi(m, \vec{p}, \theta)\,\Phi(m,-\vec{p}, \theta)
\nonumber\\
&&{\bar \Phi}(m, \vec{p}\,', \theta)\,{\bar \Phi}(m, -\vec{p}\,', \theta)\cos\left(m \Theta_{0j} p^j\right)\cos\left[m \Theta_{0j} \left(p^j - k^j\right)\right]
\eqnumber\label{mlett:b33}
\end{eqnarray}
\alphend

\noindent
where $T^S$ stands for superamplitudes.  One can convince oneself that
the effective potential arising from the amplitudes in Eq. (\ref{33})
reproduces those given in Eqs. (\ref{29}) and (\ref{30}). This is quite natural because
in the low energy regime the fermionic sector receives only contributions
from the above mentioned supergraphs. As for the low energy regime of the
bosonic sector of interest ($A + A \rightarrow A + A$ and $B + B
\rightarrow B + B$), the  only contributions are those from supergraphs
containing ${D}$ factors, which are responsible for the modifications of
the propagator (see Eqs. (\ref{13}) and (\ref{14})).

Let us  focus  on the one loop leading nonplanar contribution
($\Gamma_{NP}^{(1)}$) to the effective Lagrangian density, which is
similar to that in the noncommutative $\phi^3$ scalar field theory. It can 
be shown that up to lowest order in $g$

\begin{equation}
\label{35}
\Gamma_{NP}^{(1)}\,=\,\frac{g^2}{4}\,\frac{1}{16 \pi^2}\int_0^1 dx \int\frac{d\alpha}{\alpha}
e^{i\alpha[x(1-x)p^2 - m^2] + i\frac{p\circ p}{4\alpha} - \alpha \epsilon}\Phi(-p,\theta)\bar{\Phi}(p,\theta)
\end{equation}

\noindent
Here, $p^2 = p^2_0 - p^2_1 - p^2_2 - p^2_3$ is the square of the
norm of the Minkowskian four-vector $p$, while $p \circ p \equiv p^{\mu}
\left(\Theta^2 \right)_{\mu \nu} p^{\nu} $. Then, by  means of an analysis
similar to the one carried out in \cite{Gomis1}
for the case of the two-point function in the noncommutative scalar $\phi^3$ theory, we arrive to the conclusion that  the  unitarity constraint is violated whenever $p \circ p < 0$. Since $p \circ p < 0$ demands $\Theta_{0j} \neq 0$ \cite{Gomis1}, we conclude that time/space noncommutativity leads to a violation of unitarity.

Finally, we mention that the two-loop contribution to the nonplanar
K\"alherian effective potential ($\Phi = \mbox{constant}$) has already been found to
read \cite{Petrov2}

\begin{equation}
\label{36}
K^{(2)}_{NP}\,=\,\frac{1}{16 \pi^2}\,\int_{0}^{\infty}\,\frac{d\alpha}
{\alpha}\,\int_{0}^{1} dx\,\int_{0}^{\infty}\,dz\,e^{- i (\alpha + z)|m + 
g \Phi|^2}\,
\int \frac{d^4 k}{(2 \pi)^4}\,e^{i [\alpha x(1-x) + z]k^2 \,+\, i \frac{k\circ k}
{4\alpha}}\,.
\end{equation}

\noindent
For time/space noncommutativity, this potential develops an imaginary part and therefore leads to a violation of unitarity.

To summarize, for the NC WZ model, unitarity is indeed violated within
the relativistic regime. However, this does not preclude the existence
of a unitary low energy regime.

\acknowledgments

This work was partially supported by Funda\c c\~ao de Amparo \`a
Pesquisa do Estado de S\~ao Paulo (FAPESP) and Conselho Nacional de
Desenvolvimento Cient\'\i fico e Tecnol\'ogico (CNPq). Two of us
(H.O.G and V.O.R) also acknowledge support from PRONEX under contract
CNPq 66.2002/1998-99. A. P. has been supported by FAPESP, project
No. 00/12671-7.

\end{document}